BIOLOGICAL SCIENCES

*Correspondence to:  ichinose@anan-nct.ac.jp

# Evolution of Fairness in the Not Quite Ultimatum Game

Genki Ichinose[1,2]* and Hiroki Sayama[2]

1 Anan National College of Technology, 265 Aoki Minobayashi, Anan, Tokushima 774-0017, Japan.

2 Collective Dynamics of Complex Systems Research Group, Binghamton University, State University of New York, Binghamton, NY 13902-6000, USA.

**The Ultimatum Game (UG) is an economic game where two players (proposer and responder) decide how to split a certain amount of money. While traditional economic theories based on rational decision making predict that the proposer should make a minimal offer and the responder should accept it, human subjects tend to behave more fairly in UG. Previous studies suggested that extra information such as reputation, empathy, or spatial structure is needed for fairness to evolve in UG. Here we show that fairness can evolve without additional information if players make decisions probabilistically and may continue interactions when the offer is rejected, which we call the Not Quite Ultimatum Game (NQUG). Evolutionary simulations of NQUG showed that the probabilistic decision making contributes to the increase of proposers' offer amounts to avoid rejection, while the repetition of the game works to responders' advantage because they can wait until a good offer comes. These simple extensions greatly promote evolution of fairness in both proposers' offers and responders' acceptance thresholds.**

While traditional economic theories assumed that individuals are generally rational and self-interested, it has often been experimentally observed that real humans behave rather irrationally [1, 2]. The Ultimatum Game (UG) [3] is one such example in which player's behaviors contradict between theoretical prediction and experimental observation.

In UG, two players, a proposer and a responder, are to split a certain amount of money given to them. The proposer can make only one offer to the responder about how to split the money. If the responder accepts the offer, the money will be split between them accordingly. If the responder rejects the offer, neither player receives anything. A rational, profit-maximizing responder should accept any small yet non-zero offer, because acceptance is always better than rejection that would give him nothing. Therefore, a rational, profit-maximizing proposer who believes that the responder is also rational should claim almost the entire amount, leaving only a tiny fraction for the responder. Straightforward computer simulations of evolutionary UG have shown that these rational strategies do evolve naturally [4].

However, it has been observed in a large number of experimental studies that this is not how real humans actually behave in this game [5-10]. Instead, most proposers tend to offer a fair share to their responders, usually offering 40 to 50%, while responders frequently reject offers below 30%. It is also known that these percentages show significant variations across different countries and cultures [6, 9]. We also had conducted a scenario experiment of UG in the past, which showed that the average amount of proposers' offers was 46.7 % and the average threshold for responders' acceptance was 37.4% [10].

Several models have been proposed to explain such fair behavior observed in human subjects. The inequity aversion model assumes that humans care about not only the benefit they receive but also the equality between them and others [9], which is still based on the rationality assumption but with a different utility function. In the meantime, studies with evolutionary game theory-based models have suggested that including additional information in UG promotes the evolution of fairness. Nowak et al. (2000) showed that fairness can evolve when the information about responder's acceptance levels (reputations) are known to proposers [11]. Page and Nowak (2002) showed that empathy, i.e., players' tendency to equate offer amounts with their own acceptance thresholds, can also contribute to the evolution of fairness [12]. Moreover, it is also known that spatial structures including regular and complex networks facilitate the evolution of fairness [4, 13-16]. However, these additional assumptions are not free from problems. The inequality aversion model does not explain why such aversion exists or evolves. The reputation model does not consider the possibility that responders with reputations for high acceptance levels may lose opportunities to participate in games because proposers would tend to avoid them. The empathy model does not correctly represent actual human behaviors that typically show different offer amounts and acceptance thresholds. Spatial structures are typically assumed to be static in the previous studies, but they are the representation of social relationships of human individuals and therefore they must change quite dynamically at the time scales of behavioral evolution.

Here we propose a new evolutionary model of UG to show that fairness can evolve without additional information such as reputation, empathy, or spatial structure. Instead, our model assumes probabilistic nature in players' decision making, which is a reasonable assumption given that human decision making is often made probabilistically. Based on the idea of this probabilistic decision making, in our model, players randomly choose their interaction parameters for proposers, $p'$ (offer amount), and for responders, $q'$ (acceptance threshold), from a normal distribution with standard deviation $\sigma$ whose means are defined by their genotypes $p$ and $q$. Moreover, in the earlier studies, UG is conducted only once between two players as the name ("ultimatum") implies. In contrast, we consider the case that UG may be played more than once with some probability, $r$. This is also reasonable because negotiation in real humans often needs more than one interaction. We thus name our model the Not Quite Ultimatum Game (NQUG) model. See Methods for more details of the model.

**Results**

We first conducted systematic computer simulations starting with an initial population of players with randomly generated strategies using a preliminary version of the model that does not allow repetition of games (see Methods). Parameter settings we used were as follows: population size $n = 1000$; mutation range $\epsilon = 0.005$; length of a simulation run = 10,000 generations.

Figure 1 summarizes the simulation results for varying $\sigma$. Compared to the original UG ($\sigma = 0$), the characteristic offer amount of proposers ($\bar{p}$) increased greatly for higher values of $\sigma$. This is because the probabilistic fluctuations of responders' decisions increase the risk of rejection and the proposers will thus need to increase their offer amounts in order to avoid being rejected. In other words, fairness in offer amounts evolved just by increasing randomness in players' decision making. In this experiment, the value of $\sigma$ were assumed to be the same for both proposers and responders. We also studied cases where $\sigma$ is different between proposers and responders; see Supplementary Information (Fig. S3) for more details. However, this modification did not alter the qualitative interpretation of the results compared to cases with an identical $\sigma$ value. Therefore, we used the same $\sigma$ value for both proposers and responders hereafter.

In the meantime, Figure 1 also shows that increasing $\sigma$ does not affect evolution of the characteristic acceptance threshold ($\bar{q}$). To explore possibilities for both $\bar{p}$ and $\bar{q}$ to evolve toward higher values, we introduced an additional assumption to the preliminary model. Specifically, we introduced a new parameter, $r$, a probability for players to repeat playing the game again if an offer is rejected. We call this full version the Not Quite Ultimatum Game (NQUG) model (see Methods for details). Note that the repetition of games makes sense only in the UG with probabilistic decision making, but not in the original UG, because repeating games do not produce any different outcome in the original UG where the players' decisions are made deterministically.

It has been reported that fairness is likely to appear in repeated UG based on reciprocity [17]. Compared to that, our NQUG model is simply based on probabilistic decision making and does not require any information about the past that would be needed for reciprocity to function.

Figure 2 shows simulation results with the NQUG model, illustrating how the new parameter $r$ influences the evolution of characteristic offer amounts ($\bar{p}$, Fig. 2A) and characteristic acceptance thresholds ($\bar{q}$, Fig. 2B). All the other parameters were set to the same values as used in Fig. 1. It was observed that, when $r$ is relatively high (i.e., $r \sim 0.8$ or higher), the characteristic acceptance thresholds evolved toward a higher value when $\sigma$ is positive (Fig. 2B), because rejection may become a more attractive alternative than accepting low offers for responders if repeating the game is possible. Consequently, proposers' characteristic offer amounts also evolved toward a comparably higher value (Fig. 2A), although this trend does not continue to hold for higher $\sigma$, where responders tend to accept low offers frequently even though their thresholds are high. We also note that, when $r$ is moderate (i.e. $0.10 \leq r \leq 0.80$), the general trend of $\bar{p}$ is slightly decreasing (Fig. 2A and Fig. S6). This can be understood as follows. For $r = 0$, the proposers need to increase the offer amount in order to avoid rejection, as described in the preliminary model (Fig. 1). However, when $r$ is moderately positive, there is a reasonable chance that a game is repeated again even if the offer is rejected. This makes it possible for the proposers to act more boldly and thus reduces the offer amount a little, because even low offers may sometimes be accepted if the game can be repeated. Nevertheless, when $r$ is very high, it works for responders' benefit, as described above. In such cases, proposers have to increase their offer amount again to adapt to the responders' high demands. This is why a non-monotonic behavior is observed for $\bar{p}$ when $r$ is varied from 0 to 1 (Fig. S6).

Figure 3 summarizes the results presented in Figs. 2A and 2B in a single $\bar{p}$ - $\bar{q}$ space, showing how the evolved strategy changes when $\sigma$ and $r$ are varied. When $\sigma = 0$ (Fig. 3, bottom-left), the game is equivalent to the original UG because repetition of game play does not make any difference in this case. Therefore, the players' average strategy always converges to the most rational behavior (low $\bar{p}$, nearly zero $\bar{q}$) regardless of $r$. As $\sigma$ increases, however, the average strategy shifts rightward toward higher $\bar{p}$. For $r \sim 0.7$ or less, the increase of $\bar{q}$ is not significant, but for $r \sim 0.8$ and higher, the average strategy moves diagonally along the "perfect empathy" line ($\bar{p} = \bar{q}$), which indicates the evolution of fairness. It is also observed for higher $\sigma$ and extremely high $r$ ($r \sim 1$) that players tend to evolve to become "naysayers" (Fig. 3, top-right), always asking for more than what they would offer if they were proposers ($\bar{p} < \bar{q}$) because the rejection probability is nearly zero.

**Discussion**

In this article, we proposed NQUG as a probabilistic version of UG where player's decisions on offer amounts, acceptance thresholds and continuation of the game are all probabilistic. Simulation results demonstrated that fairness can naturally evolve for high $\sigma$ and high (but not too high) $r$. High $\sigma$ means more random fluctuations in players' decisions on offer amounts and acceptance thresholds, which elevates the risk for proposers to be rejected and thus the characteristic offer amount ($\bar{p}$) evolves to higher values. High $r$ means greater likelihood of continuation of game play if rejection occurs, which works to responders' advantage and thus the characteristic acceptance threshold ($\bar{q}$) also evolves to higher values.

It is already reported that noise at selection process also promotes the evolution of fairness [18, 19]. Such noise allows less adaptive players to survive, which contributes to the evolution of fairness, because offers based on rational, profit-maximizing behavioral principles can be rejected by such less adaptive players. Our model is different from those previous studies, because we consider the noise at the individual decision making level, instead of the noise at the evolutionary level.

Our model is still limited in several aspects. We assumed that the probability distribution used in decision making is identical for all individuals, which is obviously not the case in real humans. Furthermore, the repetition probability $r$ was assumed to be constant regardless of the number of repetitions, which may be unrealistic. The repetition probability also could be different for different pairs of individuals depending on their relationships. These model extensions are interesting issues to be addressed in the future.

Finally, we do not claim that the probabilistic nature of NQUG is the primary explanation of the fair behavior observed in human subjects in UG. Rather, the evolutionary origins of fairness are probably a combination of several mechanisms proposed and studied so far (e.g., inequality aversion, reputation, empathy, spatial structure, probabilistic nature, etc.). The main contribution of our NQUG model is to show that a simple randomness in players' decision making can, by itself, account for a substantial increase in proposer's offer amounts and responders' acceptance thresholds.

**Methods**

The preliminary model

This model describes the evolution of strategies among *n* players over time. In the preliminary model, each individual player *i* has two genetically determined traits, $p_i \in [0, 1]$ and $q_i \in [0, 1]$. The trait $p_i$ is the characteristic amount of offers the player *i* makes when he is a proposer, while the trait $q_i$ is the characteristic acceptance threshold he uses when he is a responder. The player *i*'s strategy is thus represented by a two dimensional point $(p_i, q_i)$ in a unit square [4, 11]. Let $(p_1, q_1)$ and $(p_2, q_2)$ be the strategies of the proposer and the responder, respectively. Then the proposer decides the actual offer amount $p'_1$ probabilistically by sampling it from a truncated normal distribution centered at $p_1$ within the interval [0, 1], i.e., $p'_1 \sim N(p_1, \sigma^2)$ conditional on $0 \leq p'_1 \leq 1$, where σ is a global model parameter that specifies the standard deviation of players' decision making before truncation. Similarly, the responder decides the acceptance threshold probabilistically by sampling it from a truncated normal distribution centered at $q_2$ within the interval [0, 1], i.e., $q'_2 \sim N(q_2, \sigma^2)$ conditional on $0 \leq q'_2 \leq 1$. If $p'_1 \geq q'_2$, the offer is accepted and the proposer and the responder obtain $1 - p'_1$ and $p'_1$ as their payoffs, respectively. Otherwise, the offer is rejected, and neither receives anything. In the initial setting, the strategies of players are randomly generated. The model simulates evolutionary games in the following steps: First in each game play, two players are randomly selected from the population, one as a proposer and the other as a responder. The above steps are repeated *n²* times so that each player will participate in *2n* games on average. After the *n²* game plays are completed, a new population of players is produced by repeatedly creating an offspring from a parent sampled from the current population of players using their accumulated payoffs as selection probabilities. The offspring inherits its parent's strategy $(p, q)$ with small random mutations (in $[-\epsilon, +\epsilon]$) added to the original values within the range [0, 1]. This repeats until the new population has *n* players. These steps described above constitute one generation in the simulation. Each simulation run continues for a fixed number of generations.

The NQUG model

Each time the game is repeated between two players, their roles are preserved but their decisions ($p'_1$ and $q'_2$) are sampled from the aforementioned truncated normal distributions. The repetition ends when either the offer is accepted by the responder or the players decide to discontinue the game (which occurs with probability 1−*r* after each rejection).

1. Becker, G. S. Irrational behavior and economic theory, *J. Polit. Econ.* **70**, 1-13 (1962).

The authors thank Jeffrey A. Schmidt for his comments on this work.

G.I. and H.S. designed the research. G.I. and H.S constructed the model. G.I. performed the simulation. G.I. and H.S. discussed and analyzed the results. G.I. and H.S. wrote the main manuscript text. Both authors equally contributed to this research.

Competing financial interests: The authors declare no competing financial interests.

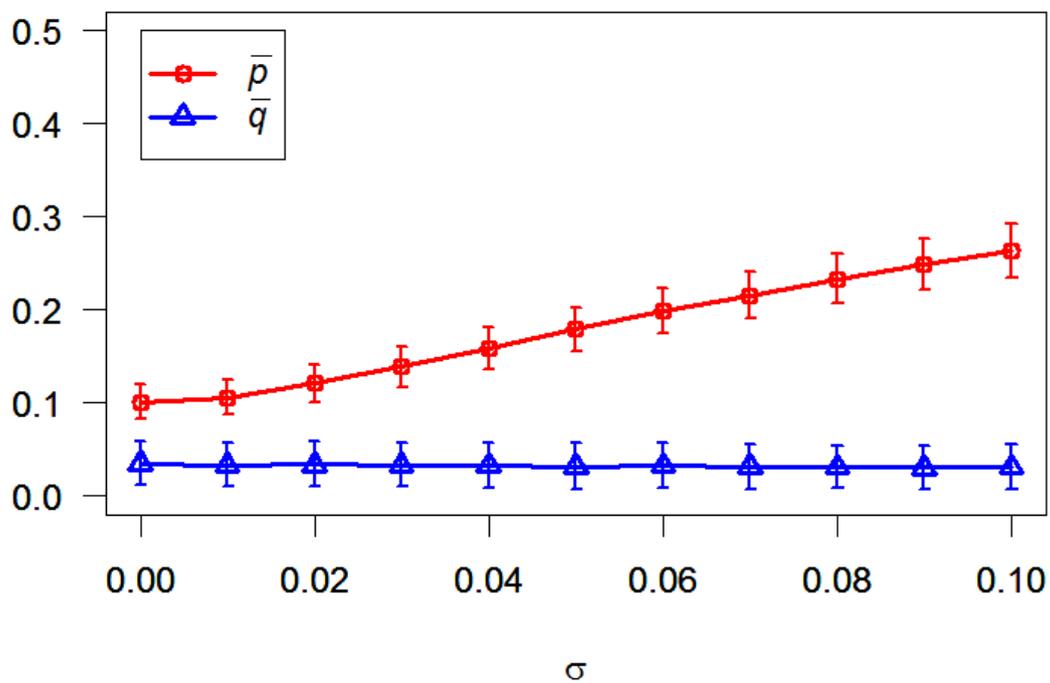

**Figure 1.** Simulation results obtained using the preliminary model. Averages of characteristic offer amounts ($\bar{p}$) and acceptance thresholds ($\bar{q}$) are plotted over varying $\sigma$. Each data point is obtained by averaging $p$ and $q$ among all the players over the last 2,000 generations of each simulation run and then averaging the measurement over 10 independent simulation runs. Error bars represent standard deviations. The average payoffs and the typical simulation runs are shown in Supplementary Information (Figs. S1 and S2, respectively).

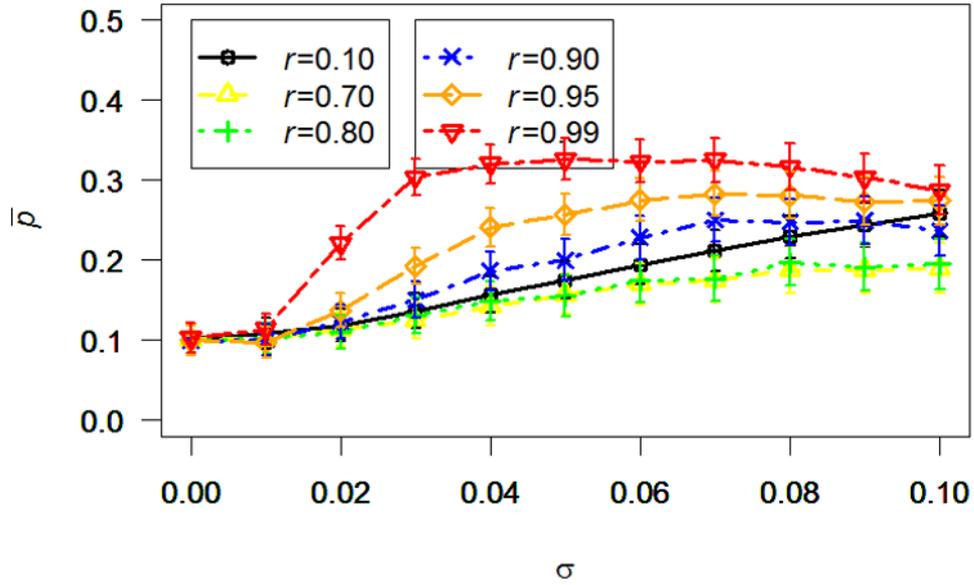

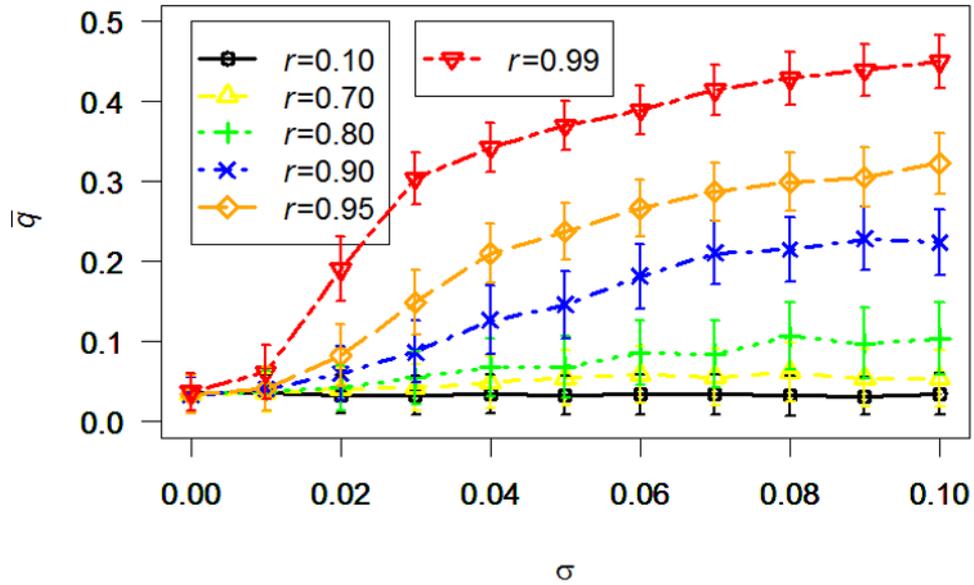

**Figure 2.** Simulation results obtained using the NQUG model. Averages of characteristic offer amounts ($\bar{p}$, **A**) and acceptance thresholds ($\bar{q}$, **B**) are plotted for varying $r$ and $\sigma$. As in Fig. 1, each data point is obtained by averaging $p$ and $q$ among all the players over the last 2,000 generations of each simulation run and then averaging the measurement over 10 independent simulation runs. Error bars represent standard deviations. The average payoffs and the typical

simulation runs are shown in Supplementary Information (Figs. S4 and S5, respectively). Both $\bar{p}$ and $\bar{q}$ evolved to high values when $r$ is high. Additional results of a more comprehensive parameter sweep experiment are provided in Supplementary Information (Fig. S6).

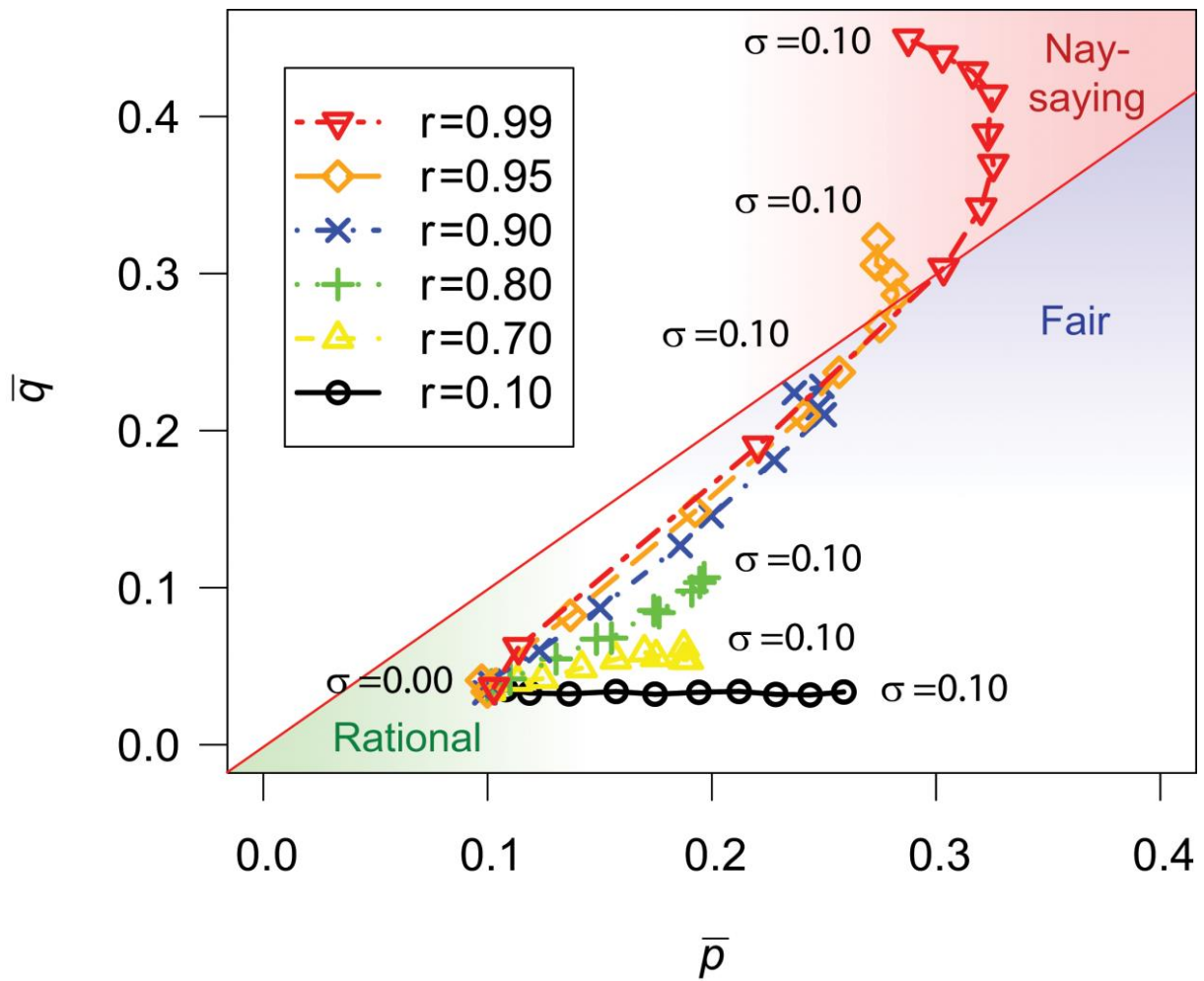

**Figure 3.** Summary of the results shown in Figs. 2A and 2B, where visualized in a $\bar{p}$ - $\bar{q}$ space. Each curve corresponds to simulation results over varying $\sigma$ while $r$ is fixed. The values of $\sigma$ are increased from 0.00 (left) to 0.10 at intervals of 0.01. Each marker represents the average of 10 independent simulation runs obtained with particular ($r$, $\sigma$) values. The diagonal line shows $\bar{p}$ = $\bar{q}$ (perfect empathy). Below the line, the bottom-left area corresponds to rational strategies predicted by traditional economics theory, while the top-right area corresponds to fair strategies observed experimentally. Above the line, the top-right area corresponds to what we call "nay-saying" behaviors, where responders tend to ask for more than what they would offer if they were proposers ($\bar{p}$ < $\bar{q}$). Such behaviors arise in our model for higher $\sigma$ and very high $r$.